\documentclass{PoS}

\title{Finite-volume effects in the evaluation of the $K_L$\,--\,$K_S$ mass difference}

\ShortTitle{Finite-volume effects in $\Delta m_K$}

\author{N.H. Christ\\
        Department of Physics, Columbia University, New York, NY 10025, USA\\
        E-mail:\email{nhc@phys.columbia.edu}}

\author{G.\,Martinelli\\
       SISSA, I-34136 Trieste, Italy \\
        E-mail: \email{martinelli@sissa.it}}

\author{\speaker{C.T.Sachrajda}\\ 
        School of Physics and Astronomy, University of Southampton, Southampton SO17 1BJ, UK\\
        E-mail: \email{cts@soton.ac.uk}}

\abstract{The RBC and UKQCD collaborations have recently proposed a procedure for computing the $K_L-K_S$ mass difference~\cite{Christ:2012se}. A necessary ingredient of this procedure is the calculation of the (non-exponential) finite-volume corrections relating the results obtained on a finite lattice to the physical values. This requires a significant extension of the techniques which were used to obtain the Lellouch-L\"uscher factor, which contains the finite-volume corrections in the evaluation of $K\to\pi\pi$ decay amplitudes. We review the status of our study of this issue and, although a complete proof is still being developed, suggest the form of these corrections for general volumes and a strategy for taking the infinite-volume limit. The general result reduces to known corrections in the special case when the volume is tuned so that there is a two-pion state degenerate with the kaon~\cite{Christ:2010gi}.
}

\FullConference{31st International Symposium on Lattice Field Theory - LATTICE 2013\\
		July 29 - August 3, 2013\\
		Mainz, Germany}

\begin{document}

\section{Introduction}
In the previous two talks by N.H.Christ~\cite{norman} and J.Yu~\cite{jianglei}, we have heard about the RBC-UKQCD programme to evaluate the long-distance contributions to $\Delta m_K\equiv m_{K_L}-m_{K_S}$, where $m_{K_{L}}$, $m_{K_S}$ are the masses of the corresponding neutral $K$-mesons. This builds on the exploratory work reported in~\cite{Christ:2012se}. To evaluate $\Delta m_K$ we need to compute the amplitude
\begin{equation} {\cal A}=\frac12\,\int_{-\infty}^\infty\,dt_1\,dt_2~T\,\langle\,\bar{K}^{\,0}\,|\,H_W(t_2)\,H_W(t_1)\,|\,K^0\rangle
\end{equation}
and extract the $K_L$-$K_S$ mass difference given by:
\begin{equation}\label{eq:DeltamKphysical}
\Delta m_K\equiv m_{K_L}-m_{K_S}=\frac{1}{2m_K}\,2{\mathcal P}\,\sum_\alpha\,\frac{\langle\bar{K}^0\,|\,{H}_W\,|\,\alpha\rangle\,\langle\alpha\,|\,{H_W}\,|\,K^0\rangle}{m_K-E_\alpha}=3.483(6)\times 10^{-12}\,\mathrm{MeV}\,,
\end{equation}
where the sum over $|\alpha\rangle$ includes an energy-momentum integral and the numerical value is the physical result.

\begin{figure}
\begin{center}
\includegraphics[width=0.8\hsize]{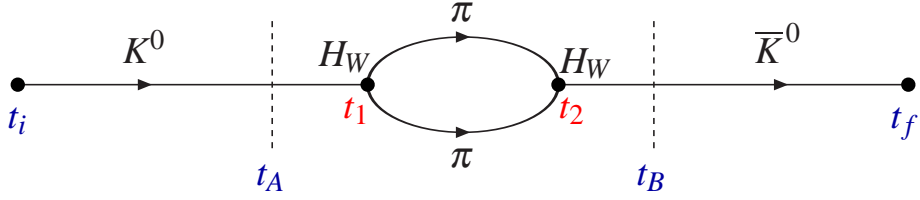}
\end{center}
\caption{Schematic diagram of the correlation function from which $\Delta m_K$ is evaluated. The $K^0$ is created at $t_i$ and the final-state $\bar{K}^0$ is annihilated at $t_f$. The times of the two insertions of the weak Hamiltonian are $t_{1,2}$ which are integrated from $t_A$ to $t_B$. The two-pion intermediate state propagating between $t_1$ and $t_2$ is drawn for illustration; the calculation includes a sum over all possible intermediate states.\label{fig:sketch}}
\end{figure}

In the lattice calculation we compute the four-point correlation function sketched in Fig.\,\ref{fig:sketch}. Defining $T=t_B-t_A+1$ (in lattice units) and integrating the times of the two insertions of the weak Hamiltonian from $t_A$ to $t_B$ we find
\begin{equation}\label{eq:c4}
C_4(t_A,t_B;t_i,t_f)=|Z_K|^2e^{-m_K(t_f-t_i)}\sum_n\,\frac{\langle\bar{K}^0\,|\,{H_W}\,|\,n\rangle\,\langle n\,|\,{H_W}\,|\,K^0\rangle}{(m_K-E_n)^2}\,\bigg\{e^{(m_K-E_n)T}-(m_K-E_n)T-1\bigg\}\,,\end{equation}
where the sum is over all allowed intermediate states and $Z_K$ is the matrix element of the interpolating operator used to create the $K^0$ and annihilate the $\bar{K}^0$ between the kaon and the vacuum. 

The method proposed in~\cite{Christ:2012se} is to identify the coefficient of $T$ from which we obtain:
\begin{equation}\label{eq:DeltamFV} \Delta m_K^{\mathrm{FV}}\equiv2 \sum_n\,\frac{\langle\bar{K}^0\,|\,{H}_W\,|\,n\rangle\,\langle n\,|\,{H}_W\,|\,K^0\rangle}{(m_K-E_n)}\,.\end{equation} The two-pion energy levels depend on the volume of the lattice.
If the volume is tuned so that there is a state, $|n_0\rangle$ say, whose energy is equal to $m_K$, $E_{n_0}=m_K$, then from the coefficient of $T$ we obtain instead
\begin{equation}2\sum_{n\neq n_0} \frac{\langle\,\bar{K}\,|\,H_W\,|\,n\rangle\,\langle\,n\,|\,H_W\,|\,{K}\rangle}
{(m_K-E_n)}\,,\end{equation} (the state $|n_0\rangle$ appears in the coefficient of $T^2$ and higher powers). 
The subject of this talk is the evaluation of FV effects necessary to relate the sums in $\Delta m_K^{\mathrm{FV}}$ to the corresponding integrals in Eq.(\ref{eq:DeltamKphysical}). This is necessary in order to obtain the physical mass difference from a realistic lattice calculation.

It will be important for the following to note that the correlation function $C_4$ itself does not have a pole as the volume is varied so that one of the two-pion energy levels approaches $m_K$ (see Eq.\,(\ref{eq:c4})). This is not the case however, when one extracts the coefficient of $T$ to obtain $\Delta m_K^\mathrm{FV}$ (see Eq.\,(\ref{eq:DeltamFV})).

The discussion here exploits the fact that only two-pion states lies below $m_K$ (contributions from three-pion states are neglected) and assumes the dominance of $s$-wave rescattering of the two pions. To simplify the discussion we consider only the dominant $I=0$ contribution; generalising the discussion to include also the $I=2$ channel is straightforward. The relation between the finite-volume sums and infinite-volume integrals for the case in which the volume has been tuned so that there is a state $|n_0\rangle$
with $E_{n_0}=m_K$ was presented by N.~Christ at the 2010 Lattice conference~\cite{Christ:2010gi}. Using degenerate perturbation theory he found
\begin{eqnarray}\Delta M_K&=&2\sum_{n\neq n_0} \frac{\langle\,\bar{K}\,|\,H_W\,|\,n\rangle\,\langle\,n\,|\,H_W\,|\,{K}\rangle}
{(M_K-E_n)}+\frac{1}{\left(\frac{\partial h}{\partial E}\right)}
\Bigg[\frac12\,\left(\frac{\partial^2 h}{\partial E^2}\right)~
|\langle n_0\,|\,H_W\,|\,K_S\,\rangle|^2
\nonumber\\ 
&&\hspace{0.75in}-\frac{\partial}{\partial E_{n_0}}\left\{\left.\frac{\partial h}{\partial E}\right|_{E=E_{n_0}}
|\langle n_0\,|\,H_W\,|\,K_S\,\rangle|^2
\right\}_{E_{n_0}=M_K}\Bigg]\,,\label{eq:normans}
\end{eqnarray}
where $h(E,L)\equiv\phi(q)+\delta(k)$, $\phi$ is a kinematic function defined in~\cite{Luscher:1990ux} and $\delta$ is the $I=0$, $s$-wave phase shift. The L\"uscher quantization condition for two-pion states (for $s$-wave dominance) is
\begin{equation}h(E,L)=n\pi\quad\mathrm{with}\quad E^2=4(m^2+k^2)~ \mathrm{and}~q=kL/2\pi\,.\end{equation}
The derivation of Eq.\,(\ref{eq:normans}) extends the Lellouch-L\"uscher  method for the derivation of the finite-volume effects in $K\to\pi\pi$ decay amplitudes to next order in degenerate perturbation theory.

The key new result of this study is given in Eq.\,(\ref{eq:deltamK2}) which suggests a natural strategy for taking the infinite-volume limit. Before presenting this result, we recall some of the salient features of finite-volume effects in the propagation and rescattering of two pions (Sec.\,\ref{sec:twopion}) and also present a one-dimensional toy example with similar properties to $\Delta m_K$ (Sec.\,\ref{sec:toy}). The derivation of Eq.\,(\ref{eq:deltamK2}) is sketched in Sec.\,\ref{sec:FVeffects} where it is also seen that Eq.\,(\ref{eq:normans}) is a special case. It should be noted however, that we are still developing a complete proof of Eq.\,(\ref{eq:deltamK2}) and so at this stage it should be considered a well-motivated hypothesis, but given its significance for the evaluation of $\Delta m_K$ we present it here.

\section{Finite-volume effects for two-pion states}\label{sec:twopion}

Before proceeding to the discussion of finite-volume effects in the evaluation of $\Delta m_K$ it is instructive to recall the derivation of the Lellouch-L\"uscher formula~\cite{Lellouch:2000pv} using the method of~\cite{Lin:2001ek}. Consider the correlation function
\begin{equation}C(t)=\int_V d^3x~~\langle\,0\,|\,J(\vec{x},t)\,J(0)\,|\,0\,\rangle=V\sum_n\,|\langle\,0\,|\,J(0)\,|\pi\pi,n\,\rangle_V|^2\,e^{-E_n t}\,,
\end{equation}
where the suffix {\footnotesize$V$} denotes finite-volume matrix elements.
As $V\to\infty$, the excitation level at fixed physics becomes large, i.e. $h(E,L)\to\infty$ and the Poisson summation formula implies that 
\begin{equation}\sum_n\,f(E_n)=\int\,dE\,\rho_V(E)\,f(E)+\sum_{l\neq 0}\int\,dE\,\rho_V(E)\,f(E)\,e^{i2\pi\, l\,h(E,L)}\,,\end{equation}
where 
\begin{equation}\rho_V(E)=\frac{dn}{dE}=\frac{q\phi^\prime(q)+k\delta^\prime(k)}{4\pi k^2}\,E\,,\end{equation}
Applying this formula to $C(t)$ we obtain:
\begin{equation} \label{eq:cf} C(t)\to V\int\,dE\,\rho_V(E)\,|\langle\,0\,|\,J(0)\,|\pi\pi,E\,\rangle_V|^2\,e^{-E t}+\textrm{exponentially small corrections.}\end{equation}

On the other hand, clustering implies that the finite-volume correlation function is equal to the infinite-volume one up to exponentially small corrections so that 
\begin{equation}\label{eq:ci}
C(t)=\frac{\pi}{2(2\pi)^3}\,\int\,\frac{dE}{E}\,e^{-Et}|\langle\,0\,|\,J(0)\,|\pi\pi,E\,\rangle|^2\,k(E)\,,
\end{equation}
where $k(E)=\sqrt{E^2/4-m^2}$.
Comparing the expressions in Eqs.(\ref{eq:cf}) and~(\ref{eq:ci}), we obtain
\begin{equation}
|\pi\pi,E\rangle
	=4\pi\sqrt{\frac{VE\rho_V(E)}{k(E)}}\,|\pi\pi,E\rangle_V\,.\end{equation}
This is the key ingredient of the Lellouch-L\"uscher formula,
(note also  the relation between the normalisations of single-particle states in infinite and finite volumes: \\ $|K(\vec{p}=0)=\sqrt{2m_KV}\,|K(\vec{p}=0)\,\rangle_V$).

\begin{figure}
\begin{center}
\includegraphics[width=0.13\hsize]{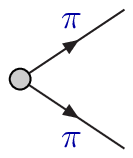}\hspace{1.5in}
\includegraphics[width=0.2\hsize]{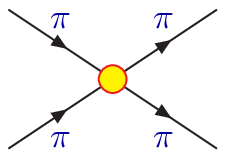}
\caption{Left - weak vertex which creates two pions from the vacuum. Right - strong vertex leading to the rescattering of two pions.\label{fig:vertices}} 
\end{center}
\end{figure}
\begin{figure}
\begin{center}
\includegraphics[width=0.65\hsize]{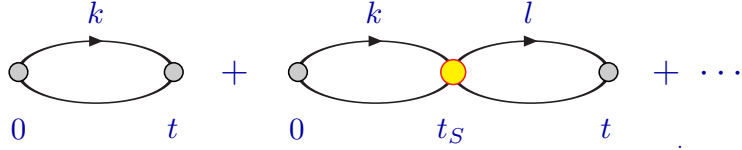}
\caption{Perturbative contributions to the correlation function.\label{fig:pert}}
\end{center}
\end{figure}
\subsection{Perturbation theory for two-pion states}
Note that although $C(t)$ has no non-exponential FV corrections, the energies and matrix elements do. It is instructive to see how this works in perturbation theory. A full calculation for $K\to\pi\pi$ decays in chiral perturbation theory was presented in~\cite{Lin:2002nq}; here we simplify the calculation while keeping the relevant points. For this illustration consider a weak vertex which can create a two-pion state from the vacuum and a strong-interaction vertex which allows the two pions to rescatter (see Fig.\,\ref{fig:vertices}).

Evaluating the diagrams in Fig.\,\ref{fig:pert}, the correlation function at zero three-momentum is found to be of the form
\begin{equation}\label{eq:oneloop}
C(t)=\frac{1}{V}\sum_{\vec{k}}\frac{1}{(2\omega_k)^2}e^{-2\omega_kt}+\frac{\hat{\lambda}}{V^2}
\sum_{\vec{k},\vec{l}}\frac{1}{(2\omega_k)^2(2\omega_l)^2}\,\frac{1}{2(\omega_l-\omega_k)}
\left\{e^{-2\omega_kt}-e^{-2\omega_lt}\right\}
\end{equation}
where $\omega_k^2=\vec{k^2}+m_\pi^2$ ($\omega_l^2=\vec{l^2}+m_\pi^2$) and $\hat\lambda$ is proportional to the strong coupling. The starting point for the approach of~\cite{Lin:2001ek} is that there are no power-like finite-volume corrections in the correlation function and this is manifested by the fact that there is no singularity at $\omega_k=\omega_l$ in the summand of the second term on the right-hand side of Eq.\,(\ref{eq:oneloop}). This does not contradict the fact that, as we know from the pioneering work of L\"uscher~\cite{Luscher:1990ux}, there are finite-volume corrections to the two-pion energies and, as we know from the Lellouch-L\"uscher formula~\cite{Lellouch:2000pv}, also to the matrix elements. To see this we combine the terms with 
$|\vec{l}\,|=|\vec{k}\,|$ in the second term on the right-hand side of (\ref{eq:oneloop}) with the lowest-order contribution to obtain:
\begin{equation}\label{eq:energyshift}
\frac{1}{V}\sum_{\vec{k}}\frac{1}{(2\omega_k)^2}e^{-2\omega_kt}+
\frac{\hat{\lambda}}{V^2}\sum_{\vec{k}}\frac{\nu_kt}{(2\omega_k)^4}\simeq
\frac{1}{V}\sum_{\vec{k}}\frac{1}{(2\omega_k)^2}e^{-(2\omega_k+\Delta E(k))\,t}\,,
\end{equation}
where $\Delta E(k)$ is the finite-volume energy shift as given by L\"uscher's formula.  
The terms with $|\vec{l}\,|\neq|\vec{k}\,|$ correctly give the Lellouch-L\"uscher formula~\cite{Lin:2002nq}.
Sum of the (power-like) finite-volume corrections to the energy and matrix elements cancel as seen in Eq.(\ref{eq:oneloop}).

The $O(\hat\lambda)$ term highlighted in Eq.\,(\ref{eq:oneloop}) comes from the integral of the time $t_S$ at which the strong-vertex in inserted from $0$ to $t$ corresponding to the propagation and rescattering of two pions which are responsible for the power-like finite-volume effects in the energies and matrix elements. The full contribution to the correlation function requires the integral over the complete range of $t_S$, the remaining terms do not contain non-exponential finite-volume effects (as demonstrated in~\cite{Lin:2002nq}).

\section{Towards understanding the finite-volume corrections to $\Delta m_K$: Some one-dimensional toy examples}\label{sec:toy}

An instructive example, with some similar features to the situation encountered in the evaluation of $\Delta m_K$  is provided by the one-dimensional formula~\cite{Testa:2005zm}\,\footnote{In Eq.\,(21) of~\cite{Testa:2005zm} there is a factor of 1/2 missing in the second term on the right-hand side of Eq.\,(\ref{eq:toy}).}:
\begin{equation} \label{eq:toy}
\frac{1}{L}\,\sum_n~  \frac{f(p_n^2)}{k^2-p_n^2} = 
{\cal P}\int_{-\infty}^\infty\,\frac{dp}{2\pi}~\frac{f(p^2)}{k^2-p^2}+\frac{f(k^2)\,\cot(kL/2)}{2k}
\,.\end{equation}
The sum on the left-hand side of Eq.\,(\ref{eq:toy}) is over the integers $n$, with $p_n=2\pi n/L$, $L$ is the length of the one-dimensional space and $k$ is an external momentum. $k$ can be considered as being analogous to the relative momentum of each pion in a $K\to\pi\pi$ decay ($m_K^2=4(m_\pi^2+k^2)$). The presence of the pole in the summand on the left-hand side, leads to non-exponential corrections to the difference of the finite-volume sum and infinite-volume integral; these corrections are given by the second term on the right-hand side.

In order to minimise the finite-volume corrections, a sensible strategy would be to tune the volume $L$ such that the $\cot(kL/2)=0$, i.e. $kL=(2n+1)\pi$ where $n$ is an integer (not to be confused with the summation variable). The infinite-volume limit would then be taken (in principle at least), by increasing $L$ while satisfying $\cot(kL/2)=0$, and for each volume only exponentially small finite-volume corrections would be encountered.

An alternative strategy might be to increase the volumes in such a way that at each step $k$ corresponds to one of the allowed momenta, $k=p_{n_0}$ say for some integer $n_0$ (which increases as the volume is increased). In that case we remove the terms $n=\pm{n_0}$ from the sum and find 
\begin{equation}\label{eq:sump}
\frac{1}{L}\sum_n\strut^\prime\frac{f(p_n^2)}{k^2-p_n^2}={\cal P}\int_{-\infty}^{\infty}\,\frac{dp}{2\pi}\,\frac{f(p^2)}{k^2-p^2}+
\frac{2f^\prime(k^2)}{L}-\frac{f(k^2)}{2Lk^2}\,,
\end{equation}
where $f^\prime(k^2)$ indicates the derivative of $f$ w.r.t. $k^2$. The $^\prime$ on the sum indicates that the terms with $n=\pm n_0$ have been removed. From the right-hand side of Eq.\,(\ref{eq:sump}) we see explicitly that  in this one-dimensional example the power-like finite-volume corrections are $O(1/L)$.

\section{Finite-volume effects in $\Delta m_K$}\label{sec:FVeffects}

Finally we return to $\Delta m_K$ and Eq.\,(\ref{eq:c4}). As already noted, the correlation function $C_4$
has no pole as one of the energies $E_n\to m_K$, and hence no non-exponential finite-volume corrections. However, the proposal in~Ref.\,\cite{Christ:2012se} is to use the $T$-behaviour of $C_4$ to extract $\Delta m_K^\mathrm{FV}$ given in Eq.\,(\ref{eq:DeltamFV}). This does have a pole as $E_n\to m_K$ and hence non-exponential finite-volume corrections. It is now convenient to make the replacement
\begin{equation}
\frac{1}{m_K-E_n}\to \frac{m_K+E_n}{4(k^2-p_n^2)}\quad\mathrm{where}\quad m_K^2\equiv4(k^2+m_\pi^2)~ \mathrm{and}~ E_n^2\equiv4(p_n^2+m_\pi^2)\,.
\end{equation}
We need to generalise the derivation of Eq.\,(\ref{eq:toy}), which starts with the relation~\cite{Kim:2005gf}
\begin{equation}\frac{1}{L}\,\sum_n~  \frac{f(p_n^2)-f(k^2)}{k^2-p_n^2}= 
\int_{-\infty}^\infty\,\frac{dp}{2\pi}~\frac{f(p^2)-f(k^2)}{k^2-p^2}\,,\end{equation}
where the summand is constructed not to have any poles. Taking 
\begin{equation}f(E_n)=2\,_V\langle\bar{K}^0\,|\,{H}_W\,|\,n\rangle_V\,_V\langle n\,|\,{H}_W\,|\,K^0\rangle_V\,
\label{eq:fdef}\end{equation}
and recalling that the quantisation condition is $h(E_n,L)=n\pi$, we obtain
\begin{equation}\label{eq:deltamK2}\sum_n\frac{f(E_n)}{m_K-E_n}={\cal P}\,\int\,dE\,\rho_V(E)\,\frac{f(E)}{m_K-E}+f(m_K)\left(\cot(h)\,\frac{dh}{dE}\right)_{\!\!m_K}.\end{equation}
The non-exponential finite-volume corrections are contained in the second term on the right-hand side. 

Given the result in Eq.\,(\ref{eq:deltamK2}), what is the best strategy for taking the infinite-volume limit? It appears to be attractive to tune the volumes at each step keeping $\cot(h(m_K,L))=0$, so that only exponentially small corrections appear. Of course, in practice the tuning will not be perfect and the cotangent term in Eq.\,(\ref{eq:deltamK2}) allows us to correct for any small mistuning. 

From Eq.\,(\ref{eq:deltamK2}), we can readily obtain the result presented by N.Christ at Lattice 2010~\cite{Christ:2010gi}. If we take the infinite-volume limit such that one of the allowed energy levels, $n_0$ say, is degenerate with the kaon, $m_K=E_{n_0}$, then Eq.\,(\ref{eq:deltamK2}) implies that 
\begin{equation}\label{eq:normansresult}
\sum_n\mbox{}^\prime\frac{f(E_n)}{m_K-E_n}={\cal P}\,\int\,dE\,\rho_V(E)\,\frac{f(E)}{m_K-E}+f^\prime(m_K)+\frac12\,
f(m_K)\,\frac{h^{\prime\prime}}{h^\prime}\,.
\end{equation}

\section{Summary and Conclusions}
Progress towards the evaluation of the $K_L$\,--\,$K_S$ mass difference~\cite{Christ:2012se,jianglei} is one of the examples of the RBC-UKQCD collaboration's programme of extending the range of physical quantities which can be evaluated in lattice simulations. In order to obtain the physical result from $\Delta m_K^\mathrm{FV}$ in Eq.\,(\ref{eq:DeltamFV}), determined from the $T$-behaviour of the correlation function, we need to correct for the finite-volume effects which are exhibited in Eqs.\,(\ref{eq:deltamK2})   and (\ref{eq:normansresult}). 

In this talk we have sketched the derivation of the powerful result in Eq.\,(\ref{eq:deltamK2}). We are currently working towards a complete proof that threshold effects in $f(E_n)$ defined in Eq.\,(\ref{eq:fdef}) do not lead to power corrections in the volume. An important consistency check is that Eq.\,(\ref{eq:deltamK2}) reproduces correctly Eq.(\ref{eq:normansresult}).

\paragraph{Acknowledgements:} NHC acknowledges partial support from the US DOE grant DE-FG02-92ER40699 and CTS from STFC grant ST/G000557/1.

\end{document}